\documentclass{article}

\usepackage[margin=1in]{geometry}
\usepackage[T1]{fontenc}
\usepackage[utf8]{inputenc}
\usepackage{microtype}
\usepackage{amsmath,amssymb}
\usepackage{booktabs}
\usepackage{enumitem}
\usepackage{graphicx}
\usepackage{xcolor}
\usepackage[numbers,sort&compress]{natbib}
\usepackage{hyperref}
\usepackage{url}
\usepackage{float}

\hypersetup{
  colorlinks=true,
  linkcolor=blue!55!black,
  citecolor=blue!55!black,
  urlcolor=blue!55!black
}

\newcommand{\devrepourl}{https://github.com/XiaoliangQi/agentic-publication-protocol-dev.app}

\title{Agentic Publication Protocol: An Attempt to Modernize Scientific Publication}

\author{
  Sirui Lu$^{1,2}$ and Xiao-Liang Qi$^{3,*}$\\[0.5em]
  $^{1}$Max-Planck-Institut f\"ur Quantenoptik, Garching, Germany\\
  $^{2}$Munich Center for Quantum Science and Technology (MCQST),\\ Schellingstrasse 4, D-80799 M\"unchen, Germany\\
  $^{3}$Leinweber Institute for Theoretical Physics, Stanford University, Stanford, CA, USA\\
  $^{*}$\texttt{xlqi@stanford.edu}
}

\date{}

\begin{document}

\maketitle

\begin{abstract}
    Scientific publication is still organized primarily around static manuscripts,
    even though much of scientific progress depends on tacit know-how: how to run
    code, reproduce figures, interpret edge cases, choose useful follow-up
    directions, and avoid failed paths. Large language model agents create an
    opportunity to publish not only knowledge, but also operational know-how in a
    form that future readers and researchers can directly use. This paper outlines
    the Agentic Publication Protocol (APP), a lightweight repository format for
    packaging a paper together with code, data, environment information,
    reproducibility instructions, and an agent-facing instruction file. APP treats a
    version-controlled repository as the publication object and uses
    \texttt{AGENTS.md} and optional skills to define a paper agent that can explain
    the work, reproduce key results when possible, and support follow-up research.
    We describe the design principles and details of the protocol, as well as the agent skills useful for publishing papers under the protocol. We also describe development tools for evaluating and improving the protocol and associated agent skills. Finally, we provide a broader discussion of the future of scientific research in the agent era. 
\end{abstract}

\section{Introduction}\label{sec: introduction}

Scientific research is a long-term, collaborative process involving the entire research community. In the current research paradigm, scientific papers are the primary medium for sharing discoveries. In addition, they also serve two other important roles. Posting and publishing a paper is the main way to establish priority and credit. The impact of papers is one of the main metrics for evaluating and rewarding researchers. Peer review, journal publication, and citation are all built around the paper as the central scholarly object.

However, the static paper format has long been criticized across different domains for a number of limitations. More specifically, it has the following problems, which cause friction and burden in the research process:
\begin{enumerate}
    \item \textbf{Limited, static representation of information.} The paper format is rigid and static. It can only target a particular group of readers. For example, a paper published in a specialized journal may be difficult to understand for readers outside the subfield, while a paper published in a general journal may emphasize a general idea that is attractive to a broader audience but correspondingly sacrifice technical precision and detailed information.
    \item \textbf{Failure to update.} A paper, once published or posted, is difficult to update. Improvements or corrections that are useful for the reader can only be made in a new version when they are significant enough. Suggestions for improvement from readers can only be manually incorporated by the authors. In the current system, the authors often lack sufficient incentives to update the paper. Especially after a paper is published in a journal, there is no way to make such updates other than publishing an erratum or a new follow-up paper. In comparison, an open-source project on GitHub can be updated easily by the authors and improved by contributions from the entire community.
    \item \textbf{Lack of reproducibility.} Many papers do not provide sufficient information to reproduce the key results, such as code, data, environment details, or instructions \cite{baker2016reproducibility,peng2011reproducible}. This makes it difficult for readers to verify the claims, build on the work, or learn from the methods. Even under journal data-sharing policies, one study of \emph{Science} papers could obtain artifacts for only 44\% of them and reproduce the main findings for only 26\% \cite{stodden2018empirical}. Even for papers with code and data, the reproduction process can be quite nontrivial, requiring the reader to spend a lot of time and effort to understand the code, set up the environment, and run the scripts. This creates a high barrier for follow-up research and slows down scientific progress.
    \item \textbf{Missing know-how.} A paper may describe the final conclusions of a project, but it often omits the operational knowledge required to build on the work: which scripts actually generate which figures, which numerical settings matter, which derivations are fragile, which data files are authoritative, and which attempted directions failed. Much of this is tacit knowledge in Polanyi's sense: we know more than we can tell \cite{polanyi2009tacit}. Such know-how is often transferred only through direct communication with the authors or through trial and error by readers.
\end{enumerate}

With the rapid development of large language models (LLMs) in recent years, LLM agents play an increasingly important role in scientific research \cite{Wang2023Scientific,qi2026agentification,bubeck2025early,lu2025theoreticalphysics}. Agents can already plan and execute experiments or simulations \cite{boiko2023coscientist,bran2024chemistrytools,liu2025vaspilot}, contribute to theoretical and mathematical discoveries \cite{gottweis2025aicoscientist,li2024modeldiscovery,brenner2026openproblem,guevara2026graviton,feng2026towards,zheng2026aicomathematician,tsoukalas2026advancing,openai2026unitdistance,alon2026remarks}, and support end-to-end research workflows from literature review and implementation to experiments and report writing \cite{lu2024aiscientist,yamada2025aiscientistv2,schmidgall2025agentlab,zhou2026paperprogram,deng2026physvec}. In a recent article by one of us \cite{qi2026agentification}, we call this process the agentification of scientific research, which we expect to substantially change research practice in many fields. This significant change is a double-edged sword. On one hand, it brings more challenges to the outdated system of scientific papers. The accelerated research process creates a more urgent need to renovate the current paper format and the peer-review and publication procedures. On the other hand, it also creates a new opportunity to modernize scientific publication. With the help of LLM agents, we can make new attempts to overcome the problems of the publication system discussed above. This is the motivation for us to propose the \href{https://github.com/LionSR/AgenticPublicationProtocol}{Agentic Publication Protocol (APP)}\footnote{\url{https://github.com/LionSR/AgenticPublicationProtocol}}.

APP is a proposal for publishing scientific work in a new format that is more dynamic and reproducible, and that preserves not only the knowledge but also the know-how of scientific work. By following the APP protocol, information about a paper is organized in a more modular way, and additional information is provided to enhance reproducibility of the key results in the paper. By including detailed instructions, the paper becomes more ready to be read and reproduced with the help of LLM agents. The goal is that an agent is delivered to the reader together with the paper contents, so that the reader can interact with a representative of the author to understand and reproduce the paper, and can therefore focus on more creative tasks such as carrying out a related research project. Having an agent represent each paper may also enable new forms of scientific collaboration, such as new ideas and follow-up directions proposed through interactions between paper agents.

The APP git repository\footnote{We use standard version-control terminology throughout the paper. A \emph{(git) repository} or \emph{repo} is a folder whose entire history of changes is tracked; a \emph{commit} is one recorded snapshot of that folder; a \emph{tag}/\emph{release} marks a specific commit as a named, citable version; and a \emph{manifest} is a small machine-readable file that records exactly which snapshot is being claimed as the publication.} contains two parts: the protocol itself, defined by \texttt{PROTOCOL.md}, and a set of agent skills that assist the authors in preparing an APP publication. The skill \texttt{publish-paper}, which uses several other skills in each step, guides the author through the procedure of organizing a paper into an APP-compliant format and publishing it as a release of a GitHub repository. However, using the skills we provide is optional. The protocol is well-defined independently of the skills, and papers compliant with the protocol can also be built in any other way, either manually or with the assistance of other agents. 

Agentification of scientific publication is a very new development, and we would like to invite the research community to join us in improving both the protocol itself and the skills that implement the protocol. Researchers in different fields likely have different emphases or domain know-how that are essential for a paper agent. Instead of a universal \texttt{publish-paper} skill, domain-specific agent skills may be developed. To provide a starting point for further development of APP, we share a \href{\devrepourl}{development repository} that contains agent skills and tools for evaluating and improving the protocol and associated skills. 

The remainder of the paper is organized as follows. In Section~\ref{sec: protocol-details}, we give a detailed overview of the protocol, including the design principles, the general structure of an APP publication, the publication workflow, and the corresponding skills. In Section~\ref{sec: dev-tools}, we discuss the development tools for improving the APP protocol and evaluate the resulting paper agents on example papers. In Section~\ref{sec: related-works}, we discuss other works related to agentic publication. Finally, in Section~\ref{sec: further-discussion}, we conclude and further discuss the future of scientific research in the agent era.

\section{Protocol Overview}\label{sec: protocol-details}

\subsection{Design Principle}

To address the problems discussed in Section~\ref{sec: introduction}, we want the APP protocol to satisfy the following principles:

\begin{enumerate}
    \item \textbf{Organized information.} Information in the publication should be organized in a modular way to minimize ambiguity and maximize usability. The repository structure
          should be modular enough that both humans and agents can locate the relevant
          material quickly. The distinction between ``ground truth" of the work and optional auxiliary material should be clear.

    \item \textbf{Enhanced reproducibility.}
          The publication should make key results as ready as possible for reproduction
          by a reader or agent. Whenever possible, the publication should include the commands, scripts,
          data, and environment needed to reproduce essential results in the paper, including figures, tables, and analytic derivations.

    \item \textbf{Version control and timestamping.}
          The publication should be a version-controlled object with a unique identifier.
          A git commit, release tag, and manifest can establish the exact file tree being
          claimed as the publication, reducing ambiguity about priority and provenance.

    \item \textbf{Know-how sharing.}
          The publication should include instructions that teach an agent how to represent
          and use the work. In addition to representing the key results of the work itself, the agent can also represent the authors' broader know-how if the authors decide to share it in the form of agent skills.
\end{enumerate}

\subsection{More Details on the Protocol}

The APP protocol is defined by \href{https://github.com/LionSR/AgenticPublicationProtocol/blob/main/PROTOCOL.md}{\texttt{PROTOCOL.md}}. It specifies the required repository structure, metadata, validation expectations, and publication object. The protocol is designed to be lightweight and flexible, allowing authors to choose how much information to include and how to organize it, as long as the required files and structure are present.
In the APP protocol, a publication is defined as a public git repository
together with a specific release. The repository is not only a place to store
the paper source. It is the publication object itself. The protocol specifies
what files should be included, which files are authoritative, how an agent
should read the repository, and how a reader can verify which snapshot is being
claimed as the published work. It does not prescribe how the repository has to
be created. The skills used to prepare, check, or release an APP publication
are useful tools, but they are separate from the protocol itself.

A complete APP publication has three layers. The first layer is the human
publication layer, which includes the manuscript, README, license, code, data,
and environment information. The second layer is the agent-facing layer,
centered on \texttt{AGENTS.md}, which tells an agent how to represent the paper
faithfully and how to navigate the repository. The third layer is the release
and verification layer, which binds the repository to a tag, commit, manifest,
and author approval. These layers together make the publication readable by
humans, usable by agents, and identifiable as a stable scholarly object.

\begin{figure}[t]
    \centering
    \includegraphics[width=0.95\linewidth]{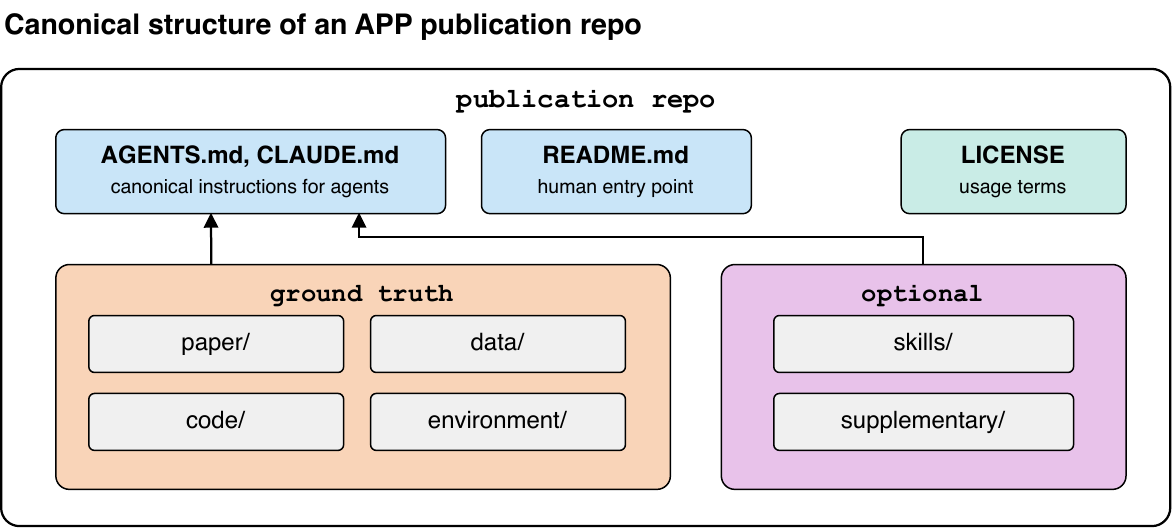}
    \caption{Repository structure of an APP publication. The repository contains
        the paper, human-facing documentation, agent-facing instructions, and the
        optional code, data, environment, supplementary material, and skills that are
        needed to make the paper usable as an agent-readable publication object.}
    \label{fig:publication-repo-structure}
\end{figure}

A typical APP repository has the following structure (see Fig. \ref{fig:publication-repo-structure}):

\begin{center}
    \begin{tabular}{ll}
        \toprule
        Path                    & Purpose                                            \\
        \midrule
        \texttt{AGENTS.md}      & Primary instructions for the paper agent           \\
        \texttt{CLAUDE.md}      & Optional pointer for Claude Code compatibility     \\
        \texttt{README.md}      & Human-facing overview and usage instructions       \\
        \texttt{LICENSE}        & Reuse terms                                        \\
        \texttt{paper/}         & Canonical manuscript and paper figures             \\
        \texttt{code/}          & Source code and reproduction scripts               \\
        \texttt{data/}          & Local data or data access documentation            \\
        \texttt{environment/}   & Dependency and runtime setup instructions          \\
        \texttt{supplementary/} & Author notes, derivations, transcripts, or context \\
        \texttt{skills/}        & Optional paper-specific agent skills               \\
        \bottomrule
    \end{tabular}
\end{center}

\texttt{AGENTS.md}, \texttt{README.md}, \texttt{LICENSE}, and
\texttt{paper/} are required for every APP publication. Other directories are
included when they are relevant to the work. For example, a theory paper with
no code or data may omit \texttt{code/} and \texttt{data/}, while a numerical
or experimental paper should include the scripts, data description, and
environment information needed for reproduction.

The \texttt{paper/} directory contains the canonical paper. This may include
the manuscript source, compiled PDF, figures, and bibliography. Exactly one
document should be identified as the canonical paper, and its format and path
should be declared in \texttt{AGENTS.md}. The \texttt{code/} directory contains
source code and scripts that belong to the publication. When figures or tables
are generated computationally, the protocol requires a figure/table
reproduction area, usually \texttt{code/figure-reproduction/}, with a
\texttt{README.md} that maps each paper figure or table to the corresponding
script, inputs, generated output, status, and any blocker. The \texttt{data/}
directory contains local datasets when they are small enough to ship with the
repository. If data are external, \texttt{data/README.md} should describe where
they come from, how to obtain them, and which scripts or figures use them.

The \texttt{environment/} directory records how to recreate the computational
environment. It may contain files such as \texttt{requirements.txt},
\texttt{environment.yml}, lockfiles, a \texttt{Dockerfile}, or other dependency
specifications. If code or nontrivial build tools are needed, the protocol
requires \texttt{environment/README.md} to state the tested platform, setup
commands, and the command prefix that an agent should use to run the code.
Installed environments and package caches should not be committed; they should
be recreated from the committed dependency files.

The central agent-facing file is \texttt{AGENTS.md}. This file contains
machine-readable frontmatter and a human-readable body. The frontmatter records
metadata such as the protocol name, protocol version, paper title, authors,
canonical paper format, publication version, domain, and optional external
references. The body should include the identity of the paper agent, a paper
summary, key results, repository structure, concrete things the agent can do,
computational requirements, and citation information. In this way,
\texttt{AGENTS.md} is not a replacement for the paper. It is an instruction
file that teaches an agent how to use the paper, code, data, and environment as
the ground truth.

The protocol also distinguishes several levels of status. A repository with an
\texttt{AGENTS.md} file is agent-readable. A repository whose
\texttt{AGENTS.md} frontmatter declares the APP protocol is an APP-structured
candidate. A verified APP publication must additionally correspond to a tagged
release with an \texttt{APP\_PUBLICATION.json} manifest.
The manifest records the repository URL, tag, commit, tree hash, validation
result, and human approval. This makes the publication a specific versioned
object rather than only a mutable folder of files.

\subsection{Publication Workflow and Skills}

To make it easier for authors to prepare an APP publication, we provide a set of agent skills in \href{https://github.com/LionSR/AgenticPublicationProtocol/tree/main/skills}{\texttt{skills/}} that guide the author through the workflow. The protocol and the publication skills are conceptually separated. The protocol specifies the required repository structure, metadata, validation
expectations, and publication object. Skills provide one convenient way to
prepare such a publication, but authors should not be required to use a
particular tool. Authors can follow the protocol manually, or use other tools or agents they prefer.

\begin{figure}[h]
    \centering
    \begin{minipage}{0.95\linewidth}
        \centering
        \textbf{(a)}\\[0.4em]
        \includegraphics[width=\linewidth]{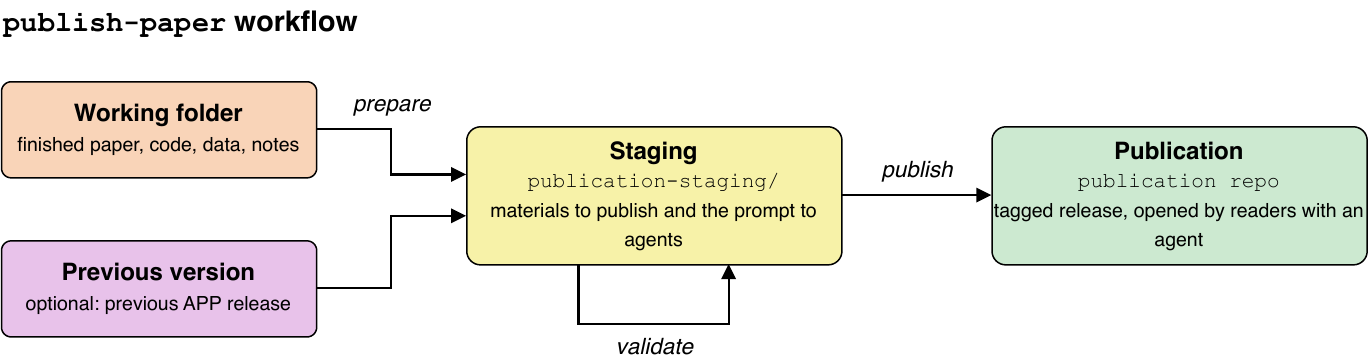}
    \end{minipage}

    \vspace{1em}

    \begin{minipage}{0.95\linewidth}
        \centering
        \textbf{(b)}\\[0.4em]
        \includegraphics[width=\linewidth]{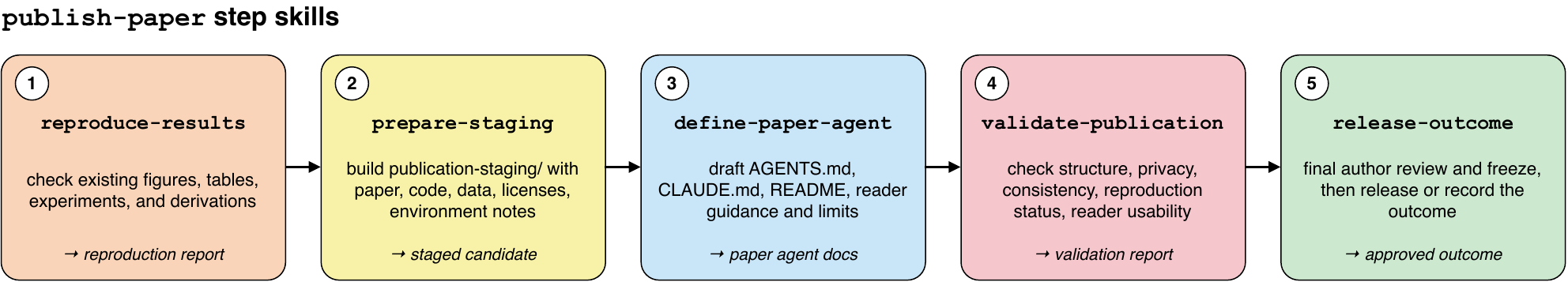}
    \end{minipage}
    \caption{Publication workflow supported by APP skills. (a) Overall workflow
        from a working paper repository to a validated APP release. (b) The
        \texttt{publish-paper} metaskill decomposes this process into step-specific
        skills that reproduce results, prepare staging, define the paper agent,
        validate the publication, and release the outcome.}
    \label{fig:publication-workflow}
\end{figure}

In the following, we provide an overview of the publication workflow defined in the \texttt{publish-paper} skill, which uses several other skills in each step. The workflow is designed to be flexible. Authors can choose to run the stages individually or run the complete workflow through the \texttt{publish-paper} metaskill. The overall workflow is illustrated in Fig. \ref{fig:publication-workflow}(a), and more detailed steps are illustrated in Fig. \ref{fig:publication-workflow}(b). The workflow starts from an original working repository, which may be a private or public repository that contains the source material of the paper. It ends with a validated APP publication release: a tagged release of the repository that corresponds to a specific commit and tree hash and has passed the validation checks.

The \texttt{publish-paper} workflow can be organized into five stages.

\begin{enumerate}
    \item \texttt{reproduce-results}.\\
          The first stage checks existing paper results before files are reorganized. The
          goal is to make the key figures, tables, experiments, and analytic results as
          reproducible as possible. Wrapper scripts should be written when useful, and
          generated figures should be visually compared with the manuscript figures. The
          outcome is a reproduction report recording which results are reproduced,
          blocked, manual-only, or inconsistent.

    \item \texttt{prepare-staging}.\\
          The second stage organizes the source material into a
          \texttt{publication-staging/} folder following the APP layout. Files are copied
          or summarized into their public locations, dependencies are documented, and
          private or irrelevant working files are excluded.

    \item \texttt{define-paper-agent}.\\
          The third stage creates the paper-agent documentation, especially
          \texttt{AGENTS.md}. This file should describe the paper's contribution, ground
          truth hierarchy, repository structure, setup instructions, reproduction map,
          limitations, and safe follow-up directions. It is helpful to keep \texttt{AGENTS.md} concise by pointing to canonical files such as \texttt{README.md}, figure-reproduction notes, data
          documentation, and environment instructions rather than repeating their
          details. Optional paper-specific skills can also be added or referred to.

    \item \texttt{validate-publication}.\\
          The fourth stage reviews the staged publication for structural completeness,
          path consistency, privacy issues, factual consistency, environment setup, and
          reader-agent usability. Validation should not act as peer review or judge
          scientific novelty. 

    \item \texttt{release-outcome}.\\
          The final stage records author approval and creates the publication object,
          such as a tagged release with a manifest. This release binds the repository
          URL, commit, tree hash, validation result, and author approval into a stable
          identifier.
\end{enumerate}

In addition to the \texttt{publish-paper} workflow and its step skills, APP
includes auxiliary skills for more targeted interactions with agent-readable
papers. The skill \texttt{load-paper} is a reader-side utility. It loads a
published paper repository, a local \texttt{publication-staging/} candidate, a
non-APP repository, or an arXiv paper into the current project so that an agent
can inspect the paper, classify its APP status, locate its documentation and
code, and help a reader consult or build on the work. When the input is a local
staging folder, the skill treats that folder as a candidate publication root, but
does not claim that it is a verified APP release.

The skill \texttt{extract-chat-context} helps authors preserve useful research
know-how from prior agent sessions. It extracts Claude Code or Codex
conversation history, summarizes the important decisions, debugging insights,
methodological choices, and dead ends, and prepares publication-safe
supplementary material such as a \texttt{know-how.md} file. This skill is useful
when interaction with agents during research is itself valuable context for future
readers. When the sessions contain recurring procedures that the research
actually validated, the skill can also propose packaging them as agent skills,
either bundled with the publication or referenced as external skills. Because chat histories can contain private notes, credentials, or
unapproved claims, the skill includes author review and privacy screening before
any extracted context is copied into \texttt{publication-staging/}.

The skill \texttt{create-paper-page} supports distribution after a paper has
been prepared. It creates a lightweight GitHub Pages project page for a paper,
including the title, authors, abstract, highlights, selected figures, links to
the paper, code, data, citation information, and an indication that the paper has
an associated agent-readable repository. The page lives under
\texttt{supplementary/} as supplementary material and is deployed to GitHub
Pages by a workflow, so it is versioned with the tagged release while the tree
keeps the protocol layout, and its links route readers to the publication
repository rather than to raw repository files. This page is not part of APP
compliance, but it provides a familiar public entry point for human readers while
linking them to the richer APP publication package.

These auxiliary skills are not required by the protocol. Rather, they illustrate
how APP repositories can become active research objects. Once the paper, code,
data, environment, and agent instructions are organized in a standard way,
different agents can reuse the same publication package for reading,
reproduction, review, presentation, and follow-up work.

\section{Agent-Assisted Development and Evaluation}\label{sec: dev-tools}

\subsection{Agentic Development Tools}
We do not intend to present APP as an already mature package of protocols and tools. Instead, our goal is to jump-start a process through which the entire research community can develop this new protocol together. To develop and improve APP, we defined several agent skills that test the performance of the protocol. In addition to sharing APP, we would also like to share a \href{\devrepourl}{development repository} with these skills, since it may be useful for researchers who are interested in contributing to the improvement of this protocol or of the development tools themselves. The goal of these agentic tools is to make the future development of APP as smooth and automatic as possible. In this section, we provide an overview of the development repository.

The published development repository is organized using the APP layout itself.
The main development materials have the following structure:

\begin{center}
    \resizebox{\textwidth}{!}{%
    \begin{tabular}{ll}
        \toprule
        Path                       & Role in development                                       \\
        \midrule
        \texttt{paper/}            & Manuscript and figures describing APP                     \\
        \texttt{code/protocol\_repo/} & Public APP protocol repository, included as a submodule \\
        \texttt{code/scripts/}     & Dataset, plotting, bibliography, and maintenance scripts  \\
        \texttt{data/example-papers/} & Source paper projects used as realistic author workspaces \\
        \texttt{data/compare-app-benchmark/} & Compact public \texttt{compare-app} summary data \\
        \texttt{skills/}           & Development skills and automated test runners             \\
        \texttt{supplementary/}    & Supplementary materials, including the project video      \\
        \texttt{working/}          & Local generated transcripts, logs, reports, and design notes, not committed in the first release \\
        \bottomrule
    \end{tabular}
    }
\end{center}

\texttt{data/example-papers/} is the public location for example source materials extracted from already published papers. We use these examples to simulate the materials (paper, code, data, notes, etc.) that authors may have at the start of the publication workflow. In the first published release, this directory is intentionally empty. Example papers can be added here as starting points for the development workflow. Some example papers are shared through our \href{https://huggingface.co/datasets/phynics/agentic-publication-protocol-dataset}{Hugging Face dataset}. A compact summary of the evaluation results discussed below, used in Table~\ref{table:compare-app} and Fig.~\ref{fig:compare-app-average}, is included under \texttt{data/compare-app-benchmark/}, while the larger transcript-level benchmark records are also hosted in the \href{https://huggingface.co/datasets/phynics/agentic-publication-protocol-dataset}{Hugging Face dataset}.

\texttt{skills/} contains the development skills used to test the protocol. The skill \texttt{simulate-publication} creates a simulated author and a publishing agent, each of which is a fresh Codex session, and runs the publication workflow in a development sandbox. The conversation history is recorded and analyzed by an evaluation agent to generate an evaluation report that points out the pros and cons of the publication procedure. The output of the publishing process is a subfolder, \texttt{data/example-papers/<example\_name>/publication-staging/}, in the example source folder. The chat history and evaluation report are saved under the local \texttt{working/} tree when the skills are run locally, which serves as a regression test for the protocol. This generated \texttt{working/} directory is not part of the first published release. After a protocol or skill change, the same source project can be tested again to see whether the workflow has become more reliable and whether previous problems have been fixed. Before each new test, the staging folder is cleared, so the agent has to prepare the staging folder from the source materials again.

After a staged publication has been created, \texttt{test-paper-agent} tests whether the resulting APP paper agent is useful to a realistic reader. It launches a simulated graduate-student reader and a paper agent inside the staged publication folder, records their conversation, and asks an evaluator agent to judge whether the paper agent answered from the APP materials, supported understanding of the paper, and gave actionable guidance for reproduction or follow-up work.

The skill \texttt{compare-app} provides a controlled comparison between an APP paper agent and a general repository-aware agent. It does not require a prior \texttt{simulate-publication} or \texttt{test-paper-agent} run; it only requires an original source workspace and a separate APP staging folder containing the paper and agent instructions. The staging folder may have been produced by \texttt{simulate-publication}, by a manual APP preparation workflow, or by any other compatible process. \texttt{compare-app} copies the paper into a paper-only reader context, generates one neutral question script from that paper context, asks the same questions to the APP paper agent in the staging folder and to the general agent in the source repository, anonymizes the two transcripts, and evaluates them without revealing which agent used the APP package. The evaluator scores four criteria: accuracy, informativeness, grounding in concrete source artifacts, and honesty about evidence and reproduction limits. Results of this evaluation on a set of example papers will be reported in Sec.~\ref{sec: dev-tools}, Table~\ref{table:compare-app}.

For these three skills, an orchestrator agent can launch multiple parallel sessions to test each example, which is helpful for speeding up the development process.

The final development skill, \texttt{find-reproducible-papers}, helps expand the example set itself. Its goal is to find papers that are useful for evaluating and improving the APP protocol. It searches for candidate arXiv papers by topic and date, then ranks them with documented signals of reproducibility. These signals include official code links, open-source licenses, repository activity, data availability statements, explicit discussion of code or reproducibility, checkable symbolic or analytic material, reproducible figures or tables, available linked artifacts such as models or datasets, and whether the work appears to require heavy compute. The shortlisted papers are then assessed more deeply by loading the paper, inspecting the code and data, and attempting to reproduce one key result. 

\subsection{Evaluation Results}

\begin{figure}[h]
    \centering
    \includegraphics[width=\linewidth]{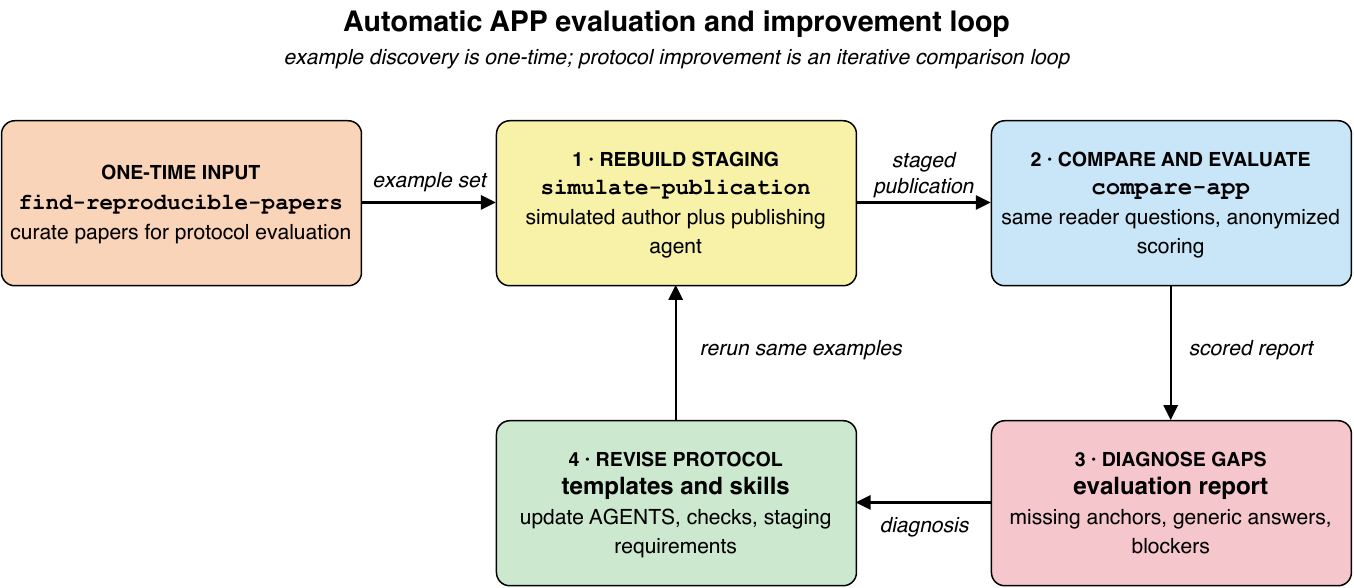}
    \caption{Automatic APP evaluation and improvement workflow. Development
        skills first help find suitable examples and rebuild APP staging folders. The
        controlled \texttt{compare-app} benchmark then probes the APP paper agent and a
        general repository-aware agent with the same reader questions. Evaluator
        reports are converted into protocol and template revisions, after which the
        same examples are rebuilt and tested again.}
    \label{fig:auto-improvement-workflow}
\end{figure}

Using the development tools described in the previous subsection, we implemented an automatic evaluation and improvement workflow, illustrated in Fig. \ref{fig:auto-improvement-workflow}. The workflow iteratively evaluates and revises the APP protocol and its agent-facing templates. More specifically, the iteration proceeds as follows:

\begin{enumerate}
    \item \textbf{Gather example papers.} A set of already published papers is curated as source workspaces under \texttt{data/example-papers/}. 
    \item \textbf{Simulate the APP publication procedure.} \texttt{simulate-publication} runs the publication workflow in a development sandbox. A simulated author and a publishing agent interact to prepare the APP staging folder \texttt{data/example-papers/<example\_name>/publication-staging/}. The full transcript and evaluator report are saved as development evidence.
    \item \textbf{Evaluate the agent performance.} Using the \texttt{compare-app} skill, we evaluate the paper agent built by \texttt{simulate-publication} against a general agent with access to the original source material. The two transcripts (with the same questions asked by the simulated reader) are anonymized and judged by a neutral evaluator according to the four criteria of accuracy, informativeness, grounding, and honesty. The evaluator reports and transcripts are converted into a short diagnosis of failure modes. This diagnosis is the basis for the next step of protocol revision.
    \item \textbf{Revise the protocol documents.} The protocol templates, staging requirements, or development skills are revised according to this diagnosis. During this procedure, we direct the agent to minimize changes to protocol documents and mainly change the template \texttt{template/AGENTS.md} and the \texttt{define-paper-agent} skill.
    \item \textbf{Iterate until satisfactory performance.} After the revision, we return to step $2$ to rebuild the publication repositories and re-evaluate them on the same example papers and question sets. The general-agent transcripts can be reused. This isolates the effect of the protocol change and avoids attributing progress to different questions or a different baseline run. The loop continues until the paper agent's performance clearly surpasses the general-agent baseline on most examples. 
\end{enumerate}

\begin{table}[H]
    \centering
    \scriptsize
    \setlength{\tabcolsep}{3pt}
    \resizebox{\textwidth}{!}{%
        \begin{tabular}{lrrrrrrrrrrl}
            \toprule
                                                                      & \multicolumn{5}{c}{APP agent}      & \multicolumn{5}{c}{General agent}  &        \\
            \cmidrule(lr){2-6}\cmidrule(lr){7-11}
            Paper                                                     & Acc. & Info. & Grnd. & Hon. & Avg. & Acc. & Info. & Grnd. & Hon. & Avg. & Result \\
            \midrule
            arXiv:2006.12469 \cite{cha2020attention}                  & 9 & 9 & 10 & 10 & 9.50 & 9 & 9 & 9 & 8 & 8.75 & APP \\
            arXiv:2412.03356 \cite{karakosta2024freespace}            & 9 & 10 & 8 & 9 & 9.00 & 9 & 8 & 9 & 8 & 8.50 & APP \\
            arXiv:2005.12702 \cite{perlin2020circuit}                 & 9 & 9 & 9 & 9 & 9.00 & 8 & 8 & 8 & 8 & 8.00 & APP \\
            arXiv:1703.10587 \cite{barghathi2017particle}             & 9 & 9 & 9 & 9 & 9.00 & 9 & 8 & 8 & 8 & 8.25 & APP \\
            arXiv:2411.03110 \cite{perezsalinas2024multiple}          & 9 & 10 & 10 & 9 & 9.50 & 9 & 9 & 9 & 9 & 9.00 & APP \\
            arXiv:2203.09758 \cite{liu2022optimal}                    & 9 & 10 & 10 & 10 & 9.75 & 9 & 8 & 9 & 9 & 8.75 & APP \\
            arXiv:2605.13271 \cite{kumar2026oam}                      & 9 & 9 & 10 & 10 & 9.50 & 9 & 9 & 9 & 9 & 9.00 & APP \\
            arXiv:2306.12711 \cite{mullerrigat2023certifying}         & 9 & 9 & 10 & 10 & 9.50 & 8 & 8 & 8 & 8 & 8.00 & APP \\
            arXiv:2310.02075 \cite{wadhwa2023learning}                & 9 & 9 & 8 & 9 & 8.75 & 9 & 8 & 8 & 9 & 8.50 & APP \\
            arXiv:2007.06989 \cite{sokolov2020emergent}               & 9 & 9 & 10 & 10 & 9.50 & 9 & 9 & 8 & 9 & 8.75 & APP \\
            arXiv:2012.01459 \cite{malz2020topological}               & 8 & 9 & 9 & 9 & 8.75 & 8 & 8 & 8 & 8 & 8.00 & APP \\
            \bottomrule
        \end{tabular}%
    }
    \caption{Public-paper subset of the fixed-harness \texttt{compare-app} evaluation. Scores are on a 10-point scale for accuracy, informativeness, grounding, and honesty; the average column is the arithmetic mean of these four aspect scores. The neutral evaluator is Codex (\texttt{gpt-5.5}, reasoning effort \texttt{xhigh}).}
    \label{table:compare-app}
\end{table}

\begin{figure}[t]
    \centering
    \includegraphics[width=0.82\linewidth]{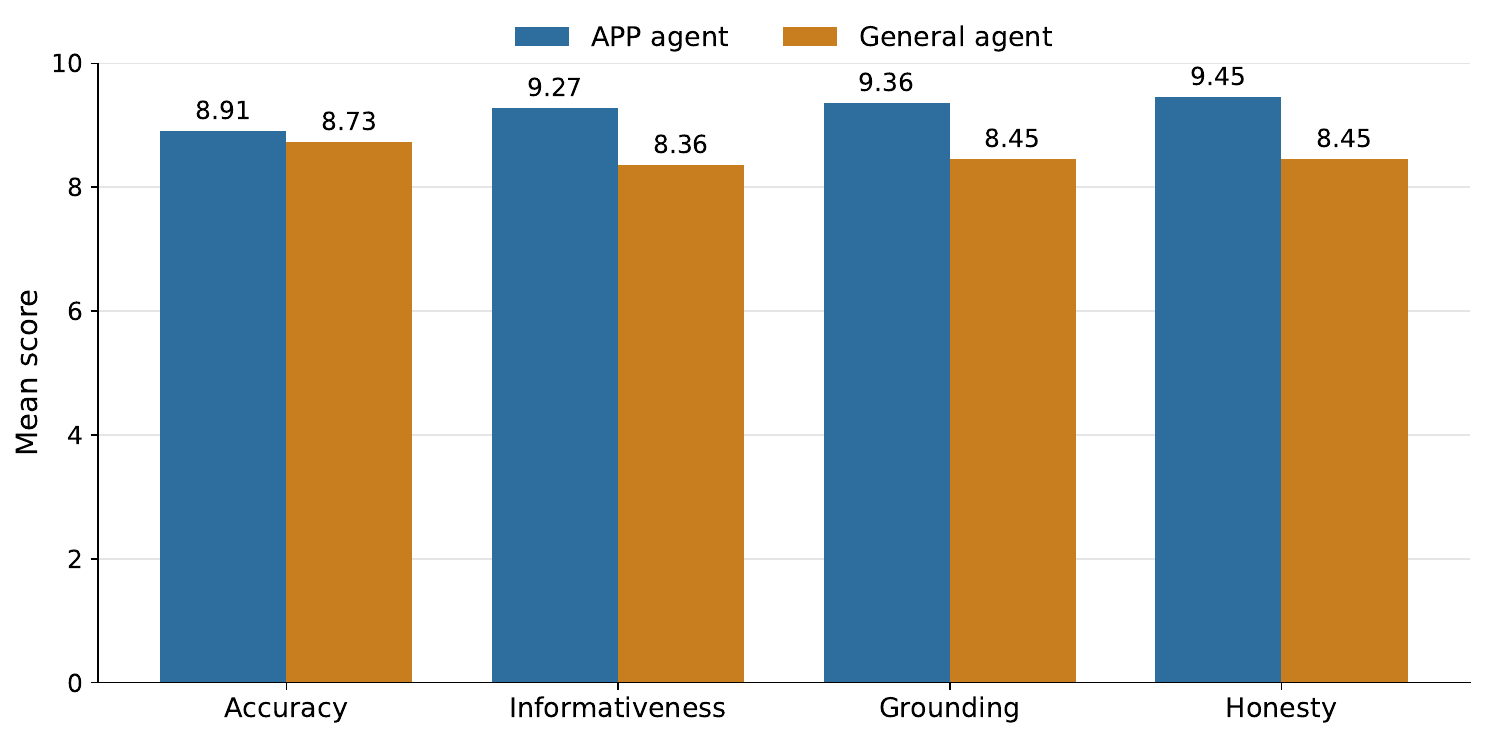}
    \caption{Average \texttt{compare-app} aspect scores over the 11 public papers.}
    \label{fig:compare-app-average}
\end{figure}

Using \texttt{find-reproducible-papers}, we built an evaluation set of 11 arXiv papers in quantum physics \cite{cha2020attention,karakosta2024freespace,perlin2020circuit,barghathi2017particle,perezsalinas2024multiple,liu2022optimal,kumar2026oam,mullerrigat2023certifying,wadhwa2023learning,sokolov2020emergent,malz2020topological}. Using 6 of these papers \cite{cha2020attention,perlin2020circuit,liu2022optimal,kumar2026oam,mullerrigat2023certifying,sokolov2020emergent} and a few other examples based on our own ongoing research work, we ran the automatic improvement loop. Initially, the paper agent had several concrete weaknesses. Early comparisons showed that the paper agent could be too distracted by protocol details and could answer from generic summaries rather than concrete source files or reproduction artifacts. After several iterations, the \texttt{define-paper-agent} skill was improved and can generate a more concise \texttt{AGENTS.md}. The performance of the improved paper agent versus the general agent is shown in Table~\ref{table:compare-app} and Fig. \ref{fig:compare-app-average}. On this data set, the APP paper agent won on all 11 papers. The mean overall score was 9.25 for the APP paper agent and 8.50 for the general agent. We also observe that performance on the ``test set'' papers is comparable to performance on the ``training set'' papers. The largest advantage of the paper agent compared with the general agent is in grounding and honesty, which is aligned with the goal of APP. 

All simulated sessions reported here were run with the OpenAI Codex CLI (versions 0.136--0.137, default model \texttt{gpt-5.5}, reasoning effort \texttt{xhigh}) in its non-interactive \texttt{codex exec} mode. The protocol itself is agent-agnostic, and the evaluation can be repeated with any agent CLI that supports scripted sessions.

Obviously, the advantage of the paper agent (built by our simulated authors) is only by a small margin, and the agent-carried evaluation may also have its own bias. The main purpose of our discussion in this section is to provide a starting point for agent-assisted development and evaluation of the APP protocol and corresponding skills. For example, a researcher interested in publishing a paper in APP can build the paper agent by herself, or with the \texttt{publish-paper} workflow, and evaluate it using the development skills provided here. We expect researchers in different fields to develop different approaches to building APP publications that can explore the further potential of this new way of sharing scientific discoveries.

\section{Related Works}\label{sec: related-works}

The limitations of the static paper discussed in Section~\ref{sec: introduction}
have motivated a long line of work that predates LLM agents. Research Objects
proposed aggregating the paper together with data, methods, and provenance into
a single shareable scholarly unit \cite{bechhofer2013linked}; the FAIR
principles made machine-actionability a community standard for research data
\cite{wilkinson2016fair}; and tools such as Binder build interactive
computational environments directly from the configuration files of a public
repository \cite{jupyter2018binder}. Many venues and platforms have also
institutionalized reproducibility checklists \cite{pineau2021improving},
artifact evaluation \cite{acm2020badging}, and paper--code links, the latter
popularized by Papers with Code and now hosted by Hugging Face
\cite{huggingfacepapers}. APP
inherits the goals of this infrastructure but changes the consumer: the
artifacts are packaged with explicit instructions so that an agent can operate
the publication on the reader's behalf, in a decentralized form that relies
only on git and public hosting conventions.

A closely related proposal is the Agent-Native Research Artifact (ARA)
\cite{liu2026ara}. ARA and APP share the view that a narrative paper plus an
unstructured code repository is insufficient for agents to faithfully
understand, reproduce, and extend research. ARA responds by proposing a typed,
layered research object whose narrative paper is a compiled view of richer
machine-readable content, including scientific logic, executable components,
exploration traces, and claim-level evidence. APP takes a lighter operating
point: it is a publication packaging and release protocol, not a universal
ontology for research content. It keeps the manuscript authoritative, uses a
familiar repository layout, and adds \texttt{AGENTS.md} plus a release manifest
so that the publication is agent-readable, auditable, and verifiable. The two
approaches are therefore complementary: ARA-like claim-evidence maps and
research traces could serve as optional deep-knowledge layers inside an APP
release, while APP provides the stable, field-neutral public release object.

The vision of the paper agent as a new interface to a publication \cite{qi2026agentification} is shared across an emerging line of work. Paper2Agent
\cite{miao2025paper2agent} realizes one version of this vision by converting an
existing paper and its codebase into a Model Context Protocol (MCP) server
\cite{anthropic2024mcp} so that a reader's agent can invoke the paper's methods
as tools. Paper2Agent is a third-party method for turning existing papers into
agentic artifacts, whereas APP is an author-side protocol for releasing such
artifacts at publication time. The two are compatible: a Paper2Agent-style tool
can consume an APP repository as a more reliable starting point. In a
complementary direction, Paper2Code reconstructs
an executable code repository from the paper text alone \cite{seo2026paper2code}. 
APP reduces the need for such reconstruction by shipping the authoritative
code, environment, and reproduction map with the publication.

More broadly, recent work on LLMs in science shows that agents are beginning to
participate across the whole research cycle, from generating ideas and running
experiments to writing papers and assisting peer review
\cite{lu2024aiscientist,bianchi2026agents4science}. LLM-generated manuscript
feedback can overlap with human expert reviews \cite{liang2024llmfeedback}, and
AI-assisted reviewing is already visible in major conferences
\cite{latona2024aireviewlottery}. At the same time, empirical studies find
growing LLM use in scientific writing and changes in manuscript production,
citation behavior, and literature discovery
\cite{liang2025usage,kusumegi2025production}, while commentary emphasizes both
the promise of AI-assisted communication and risks to depth, accountability, and
trust \cite{bertolo2024publishing}. These developments expose practical risks
such as hallucinated claims or references, fragile implementation details, prompt
injection, sycophantic review, and unclear human--AI contribution boundaries. Major venues and journals have responded with policies requiring
disclosure of LLM use and keeping humans responsible for published content
\cite{icml2023llmpolicy,nature2023chatgpt,thorp2023chatgpt,neurips2025llmpolicy,iclr2026llmpolicy}. Across different stages of the research cycle, a new consensus needs to be built about how to use AI tools responsibly to benefit scientific research. APP is an effort to build such a consensus at the stage of initial publication. Similar protocols may emerge in other stages of the research cycle, and it would be interesting to study how they interact with each other. 

\section{Conclusion and Further Discussion}\label{sec: further-discussion}

In conclusion, in this paper we proposed the Agentic Publication Protocol as a
lightweight format for making scientific publications more organized, reproducible, informative, and interactive. Instead of static knowledge objects such as manuscripts and code, an APP publication is a ``live'' representation of a research work that interacts with readers. By organizing information in the paper in a more modular way and providing more explicit reproduction instructions, an APP publication aims to reduce the burden of understanding and reproducing a research work. In addition, APP provides a way for authors to share important know-how that is usually not included in traditional publication. The goal of APP is to make scientific publications more agentic while also preserving their traditional roles as records of scientific discovery and carriers of credit and priority. 

In addition to proposing the APP protocol itself, this paper also provides two layers of agentic tools related to the protocol. The first layer consists of agent skills for publishing an APP paper, which guide authors through the workflow of preparing an APP publication. This layer also includes auxiliary skills that are useful for authors or readers. The second layer consists of agentic development tools, which help evaluate and improve the protocol and the skills in the protocol repository. We hope these tools can make it easier for researchers to adopt and contribute to the development of this new protocol. Our small-scale test shows that the
APP agents have an advantage in grounding and honesty compared with a general agent with the same LLM. 

In the final part of this paper, we zoom out to consider how the growing use of LLM agents may reshape scientific research more broadly. APP is an initial step toward a broader transformation of the research workflow. Scientific research can be viewed as a self-organized network in
which each node has inputs, processing, and outputs:
\begin{itemize}[leftmargin=2em]
    \item \textbf{Input:} gathering information from the literature, connecting
          related works, and importing prior methods.
    \item \textbf{Processing:} producing new research through analytic work,
          numerical simulations, experiments, or theory building.
    \item \textbf{Output:} sharing results through writing, posting, peer review,
          journal publication, lectures, discussion, video, audio, and other
          forms of interaction.
\end{itemize}

\begin{figure}[t]
    \centering
    \includegraphics[width=0.9\linewidth]{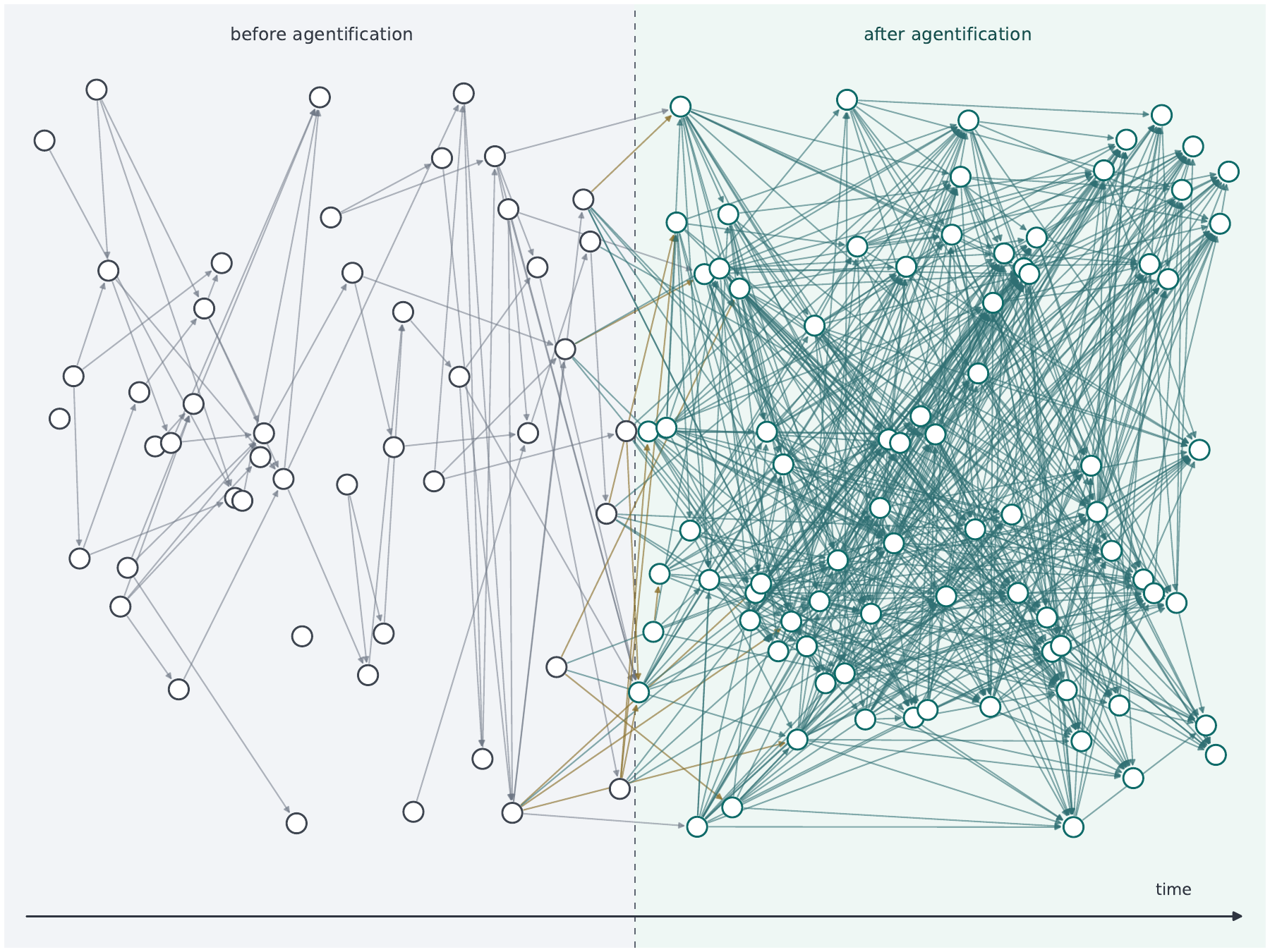}
    \caption{Schematic illustration of the qualitative change in the research
        network introduced by agentification. Each circular node represents a
        research work, and directed links point forward in time. }
    \label{fig:research-network}
\end{figure}

AI agents are bringing significant change to all three stages of research work. They can increase the degree of input by helping researchers search, connect, and operationalize prior work. They can accelerate processing by assisting with brainstorming, numerical simulations, analytic derivations, and even carrying out experiments. They can expand output by making publications interactive and usable as computational objects rather than static documents, which is the focus of APP in the current work. With AI agents, information transmission between research works becomes more efficient, informative, and interactive. The research network previously formed by human researchers is now gradually becoming a \textbf{self-organized human-AI network}. We expect such change to have the potential to drive a ``phase transition" in the dynamics of the research network, where the pace of innovation is no longer limited by the communication and collaboration burden of individual researchers. This transition is illustrated in Fig. \ref{fig:research-network}. It would be interesting to look for more precise and quantitative characterizations of such a phase transition and its implications. We leave this for future work. 

\section*{Acknowledgments}
We would like to thank Abhinava Chatterjee, Chaoxing Liu, and Quansheng Wu for testing the protocol and providing helpful feedback. In writing this paper, we used AI agents including Codex, Claude Code, TeXRA.ai for drafting part of the content, editing the language, and generating the figures. This paper is supported by the Simons Foundation (XLQ) and by the Deutsche Forschungsgemeinschaft (DFG, German Research Foundation) under Germany's Excellence Strategy - EXC-2111 - 390814868 (SL).

\section*{Data and Code Availability}
The data and code associated with this work include the
\href{https://github.com/LionSR/AgenticPublicationProtocol}{APP protocol repository},
the \href{https://github.com/XiaoliangQi/agentic-publication-protocol-dev.app}{development repository}
(which is also the APP paper repository for this work), and the example papers
and \texttt{compare-app} benchmark data in
\href{https://huggingface.co/datasets/phynics/agentic-publication-protocol-dataset}{our Hugging Face dataset}.

\bibliographystyle{unsrtnat}
\bibliography{refs}

\end{document}